\documentclass[slac_one]{revtex4}
\usepackage{subfigure}
\usepackage{hyperref}
\usepackage{graphicx}
\usepackage{fancyhdr}
\begin{document}

\title{{\small{Hadron Collider Physics Symposium (HCP2008),
Galena, Illinois, USA}}\\ 
\vspace{12pt}
Mixing and CP Violation at the Tevatron} 

%

\author{G. Brooijmans, on behalf of the CDF and D0 Collaborations}
\affiliation{Columbia University, New York, NY, USA}

\begin{abstract}
Measurements of meson mixing and CP violation parameters obtained
by the CDF and D0 experiments at the Fermilab Tevatron are presented.
These include results on $B_s$ and $D$ meson mixing, and searches for
CP violation in the decay $B^+ \to J/\psi K^+$, in mixing through 
semileptonic $B_s$ meson decays, and in the interference between mixing
and decay in the process $B_s \to J/\psi \phi$.
\end{abstract}

\maketitle

\thispagestyle{fancy}


\section{INTRODUCTION \label{sec:intro}}

The Tevatron is a heavy flavor factory in which charm and beauty hadrons of
all types are produced at high rates.  Even though the hadron collider environment
increases the complexity of events, both the CDF and D0 experiments are making
critical measurements of meson mixing and CP violation parameters.  CDF exploits its 
large level one trigger bandwidth and level two impact parameter trigger to
collect large samples of long-lived particles, and uses $dE/dx$ and time-of-flight
measurements to suppress backgrounds.  D0 benefits from its large acceptance 
for muons to collect large semileptonic samples.  In this paper, a number of unique 
results are presented: the measurements of $B_s$ and $D$ meson mixing parameters, 
and searches for
CP violation in the decay $B^+ \to J/\psi K^+$, in mixing through 
semileptonic $B_s$ meson decays, and in the interference between mixing
and decay in the process $B_s \to J/\psi \phi$.  Throughout this paper,
charge conjugated modes are implicitly assumed.

\section{MESON MIXING \label{sec:mixing}} 
Because the neutral meson mass eigenstates can be a linear combination of the weak eigenstates,
they can mix.  Taking for example $B$ mesons, we can write\cite{Yao:2006px}
\begin{equation}
|B_{L,H}\rangle = p|B^0_f\rangle + q|\bar{B}^0_f\rangle,
\end{equation}
where $B_{H,L}$ are the mass eigenstates and $B^0_f, \bar{B}^0_f$ are the weak
(or flavor) eigenstates.  The mass eigenstates then have masses 
$m_{H,L} = M \pm \frac{\Delta m}{2}$ and lifetimes 
$\Gamma_{H,L} = \Gamma \pm \frac{\Delta \Gamma}{2}$.  Because of the mass
difference, a state produced as $B^0 (\bar{B}^0)$ can oscillate between 
$B^0$ and $\bar{B}^0$ and decay as $\bar{B}^0 (B^0)$, with a characteristic
oscillation frequency $\Delta m$.  In addition to the 
variables $\Delta m$ and $\Delta \Gamma$, it is usual to define
\begin{equation}
x = \frac{\Delta m}{\Gamma},\ y = \frac{\Delta \Gamma}{2\Gamma}
\end{equation}
and
\begin{equation}
x' = x \cos\delta + y \sin\delta, y' = -x\sin\delta + y \cos\delta,
\end{equation}
which are useful when treating integral oscillation probabilities.  Here
$\delta$ is a strong phase.

\subsection{$B_s$ Meson Mixing \label{subsec:bsmix}}
From the point of view of Feynman diagrams, $B$ meson mixing is dominated by 
box diagrams with $W$ bosons and up-type quarks in the internal lines.  Since
$V_{tb}$ is large, diagrams with top quarks are dominant and the oscillation 
frequency is proportional to $m^2_{top}/m^2_W$.  The high oscillation frequency of $B$
mesons was in fact the first experimental indication of a very large top 
quark mass~\cite{Albajar:1986it,Albrecht:1987dr}, and this process is in general
sensitive to loop-level contributions from new heavy particles.  Since 
$V_{ts} > V_{td}$, $B_s$ meson oscillations are faster than $B_d$ oscillations
and the frequency is correspondingly more difficult to measure.

The analysis strategy for the measurement of the $B_s$ meson mixing frequency,
and most of the other measurements described here, consists in two main parts.
On the ``reconstructed side'', we have the $B_s$ meson under study: its
flavor at decay time is given by the charges of the decay products, and it is then
sufficient to measure its momentum and transverse decay length to determine 
its proper lifetime.  To tag the quark flavor at production time, either the 
``opposite side'' $b$-hadron is used, exploiting the fact that at the Tevatron
the vast majority of $b$ quarks are produced in pairs, or the ``same side'' hadron
from fragmentation is identified.  In the case of $B_s$ meson production, this is
likely to be a $K$ meson.

The $B_s$ meson reconstruction proceeds through the reconstruction of the decay
chain $B_s \to D_s X$, where $X$ can be leptons, or one or three pions.  Multiple 
$D_s$ decay channels are considered.  While the semi-leptonic channels allow for a 
higher trigger efficiency, they suffer from higher backgrounds and a significantly 
worse $B_s$ meson momentum determination due to the escaping neutrino.
\begin{figure*}[t]
\centering
\subfigure[]{
\includegraphics[width=0.35\textwidth]{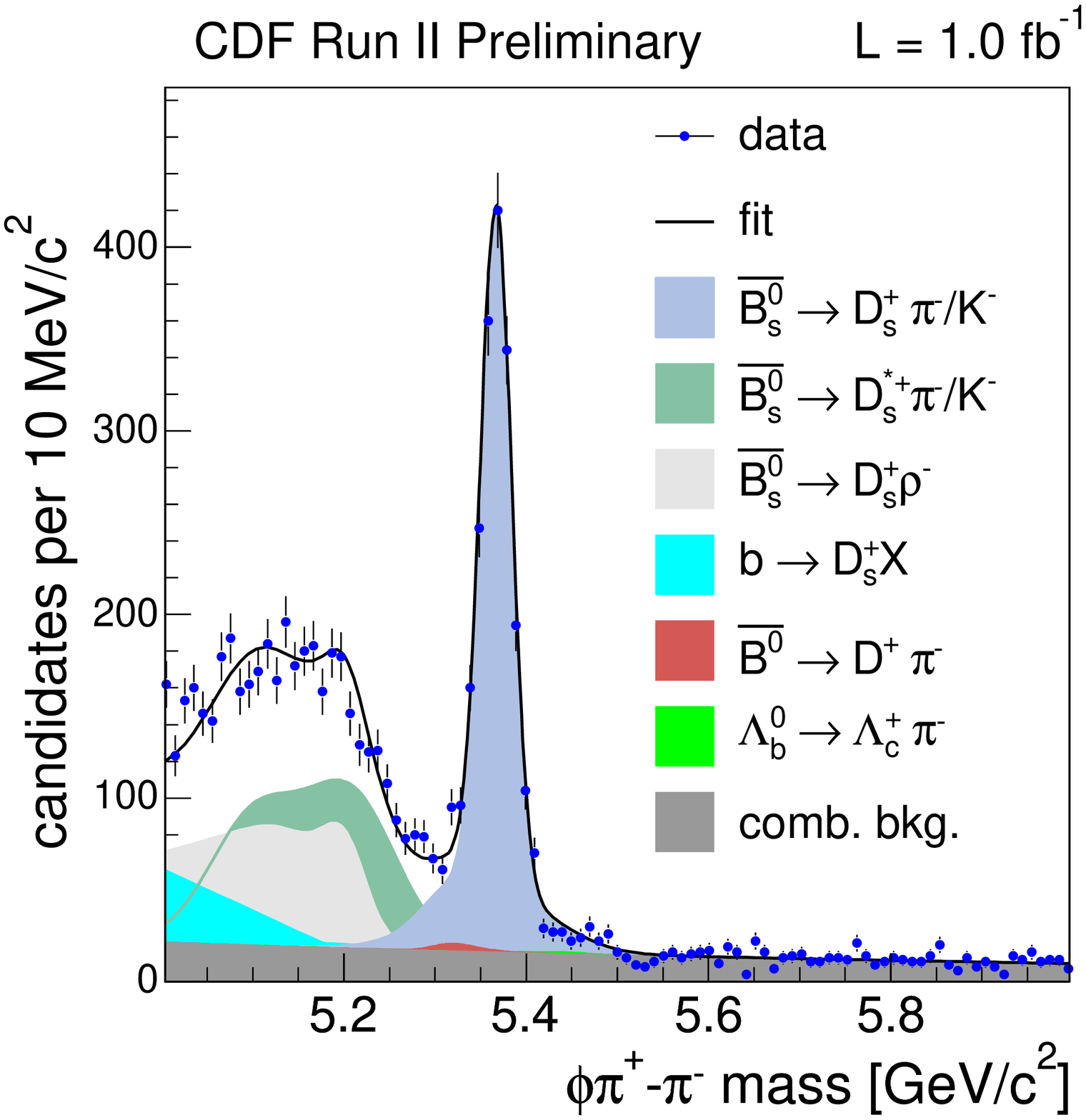}
}
\subfigure[]{
\includegraphics[width=0.35\textwidth]{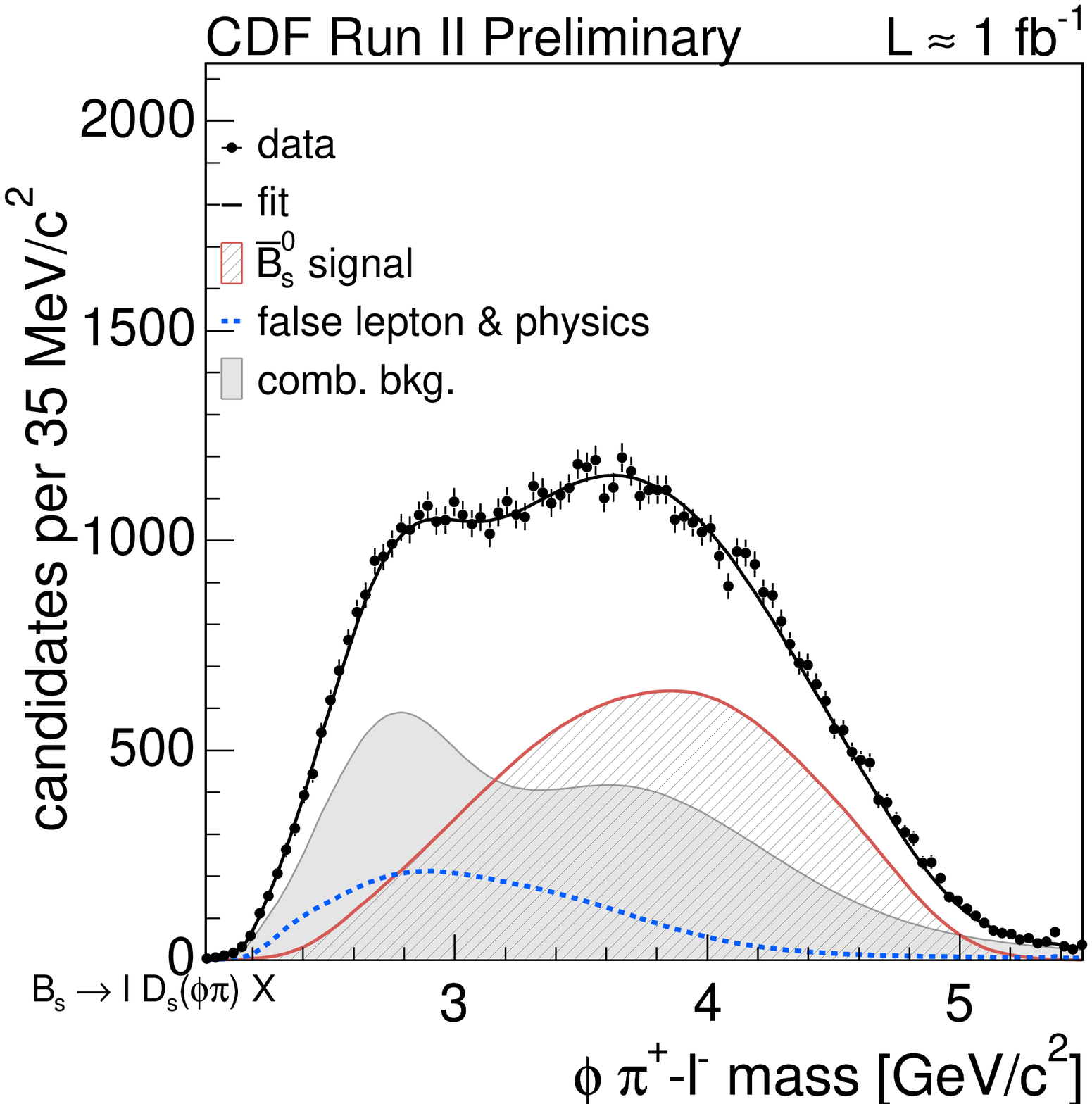}
}
\caption{Reconstructed $B_s$ candidate mass in a hadronic (left) and semi-leptonic
(right) channel at CDF.} \label{fig:bmixmass}
\end{figure*}
This is seen in Fig.~\ref{fig:bmixmass} where the reconstructed $B_s$ candidate mass
is shown for a hadronic and semi-leptonic channel used in CDF.  

The flavor of the $B_s$ meson at production is determined using 
opposite-side tagging variables that include lepton charge,
jet charge and secondary vertex charge, and same-side tagging based on 
the charge of a nearby hadron.   
Both experiments use multivariate techniques to combine the various opposite- and
same-side tagging variables:  
CDF uses a neural network and D0 uses a likelihood.
To verify the performance (efficiency and dilution) of the flavor tagging algorithm,
the $B_d$ oscillation frequency is measured~\cite{cdftag,Abazov:2006qp} 
and compared to the world average.

\begin{figure*}[t]
\centering
\subfigure[]{
\includegraphics[width=0.35\textwidth]{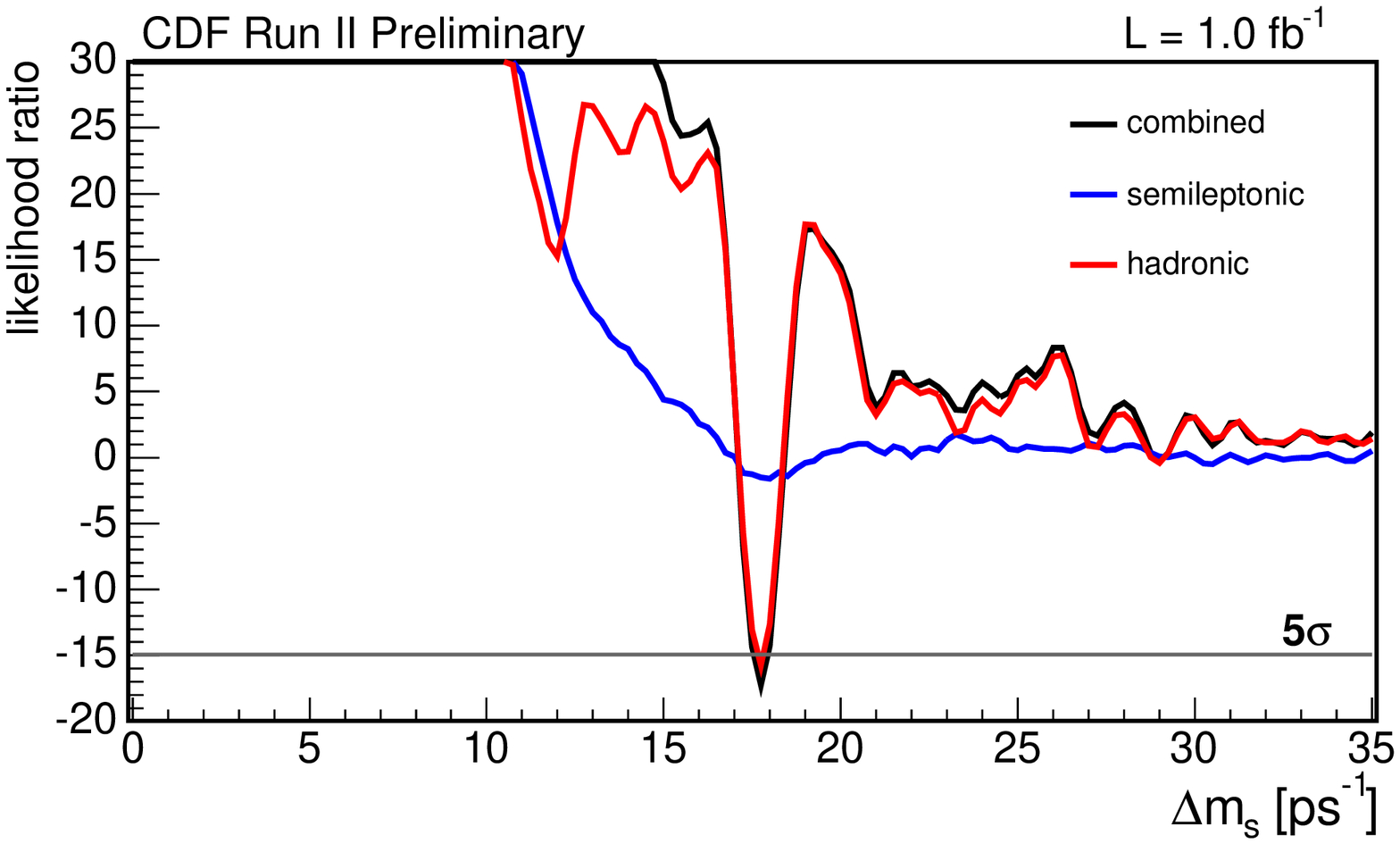}
}
\subfigure[]{
\includegraphics[width=0.35\textwidth]{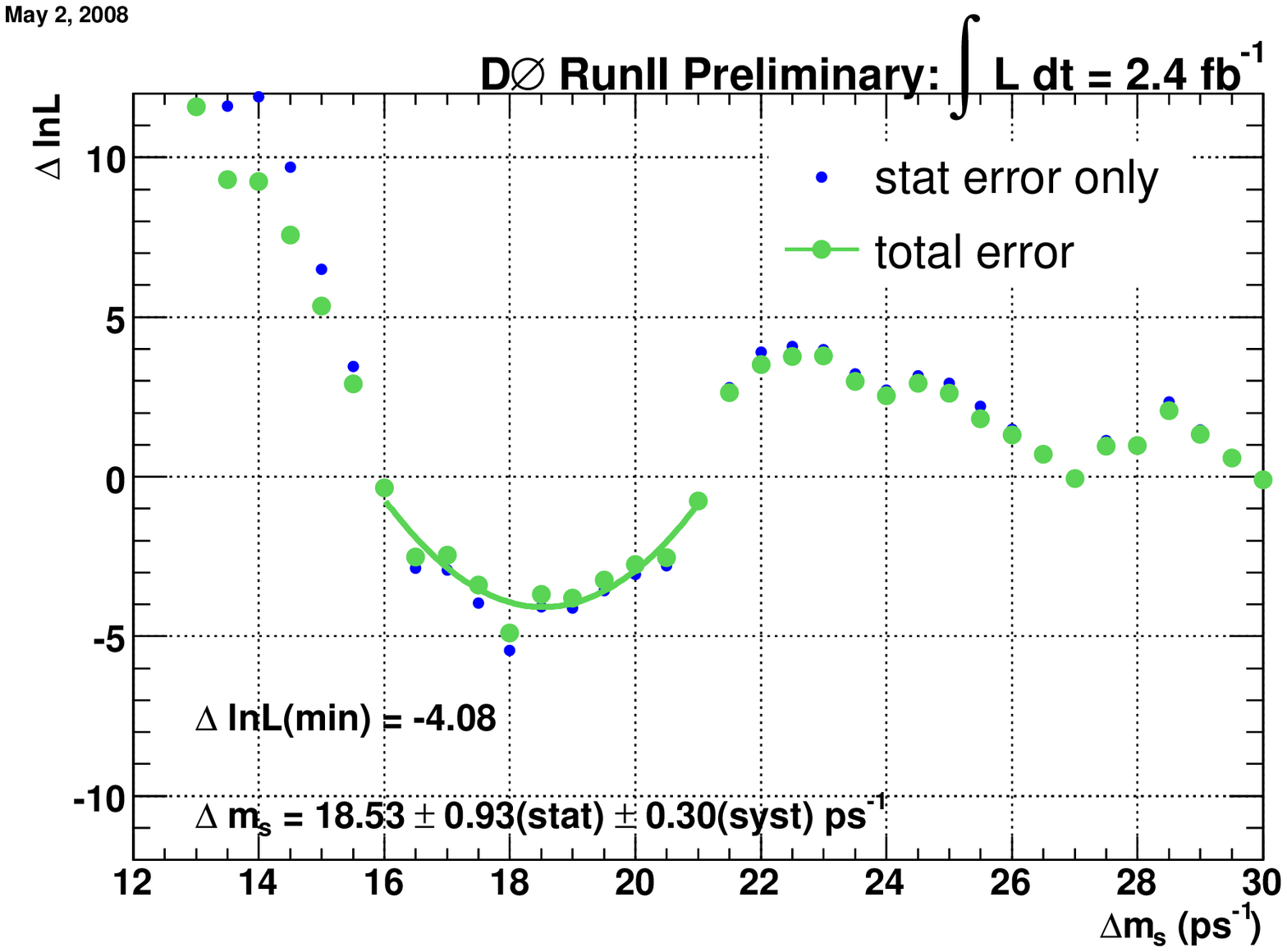}
}
\caption{Measurement of the $B_s$ meson mixing frequency.  The result is shown 
in the form of a scan over the value of the likelihood function for CDF (left)
and D0 (right).} \label{fig:bmixf}
\end{figure*}

To measure the oscillation frequency, it is necessary to extract all the 
information on an event-by-event basis, giving events weights in the 
extraction of the result that are 
proportional to their probability of originating from a certain source.  
In both experiments, an unbinned maximum
likelihood fit is used.
Figure~\ref{fig:bmixf} shows the measured $B_s$ meson oscillation frequency in 
the form of a scan over the likelihood function output.
CDF has significantly better sensitivity thanks to much larger level one trigger
bandwidth, a secondary vertex trigger at level two, and particle identification
based on $dE/dx$ in the tracker and a time-of-flight detector.  This allows
CDF to exploit the excellent resolution of the hadronic channels; they 
measure~\cite{Abulencia:2006ze}
$\Delta m_s = 17.77 \pm 0.10 (stat) \pm 0.07 (syst) ps^{-1}$ with a signal 
larger than 5$\sigma$.  In D0, where the signal is dominated by the 
semileptonic channels, the signal has a significance of 3$\sigma$ and
the measured value is~\cite{Abazov:2006dm,d0bmix} 
$\Delta m_s = 18.53 \pm 0.93 (stat) \pm 0.30 (syst) ps^{-1}$.

\subsection{$D$ Meson Mixing \label{subsec:dmix}}
\begin{figure*}[t]
\centering
\subfigure[]{
\includegraphics[width=0.25\textwidth]{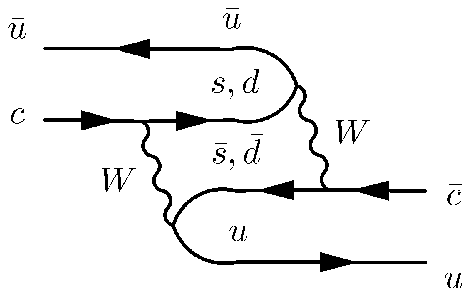}
}
\subfigure[]{
\includegraphics[width=0.25\textwidth]{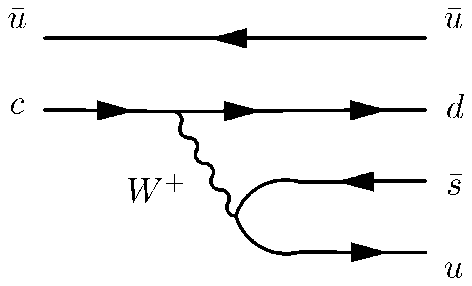}
}
\caption{Left: dominant ``long range'' contribution to $D$ meson mixing.  The process
goes through an intermediate virtual $\pi$ or $K$ meson.  Right: doubly 
cabibbo-suppressed decay.} \label{fig:dmixl}
\end{figure*}

As opposed to $B$ meson oscillations, $D$ mesons are expected to oscillate very slowly
in the standard model.  This is because the box diagram, with $\Delta C = 2$, is very
small due to a combination of CKM and GIM suppressions.  The dominant contribution 
comes from the ``long range'' ($2 \times \Delta C=1$) diagram shown in 
Fig.~\ref{fig:dmixl}, and is too small to make the oscillation observable before decay.
The effect of mixing can however possibly be seen through its interference in
the doubly Cabibbo-suppressed $D^0 \to K^+ \pi^-$ decay (right panel of 
Fig.~\ref{fig:dmixl}).

\begin{figure*}[t]
\centering
\subfigure[]{
\includegraphics[width=0.35\textwidth]{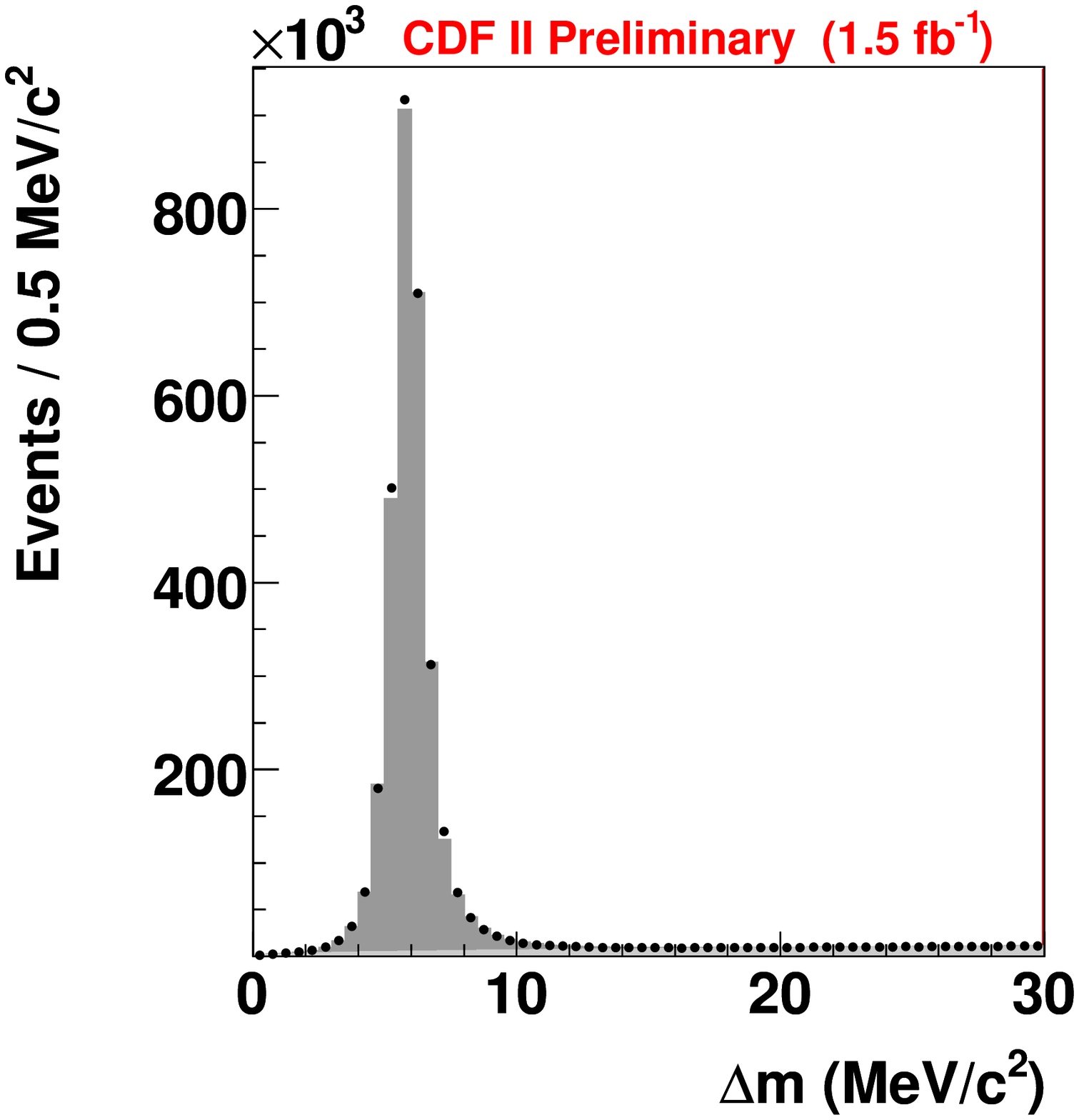}
}
\subfigure[]{
\includegraphics[width=0.35\textwidth]{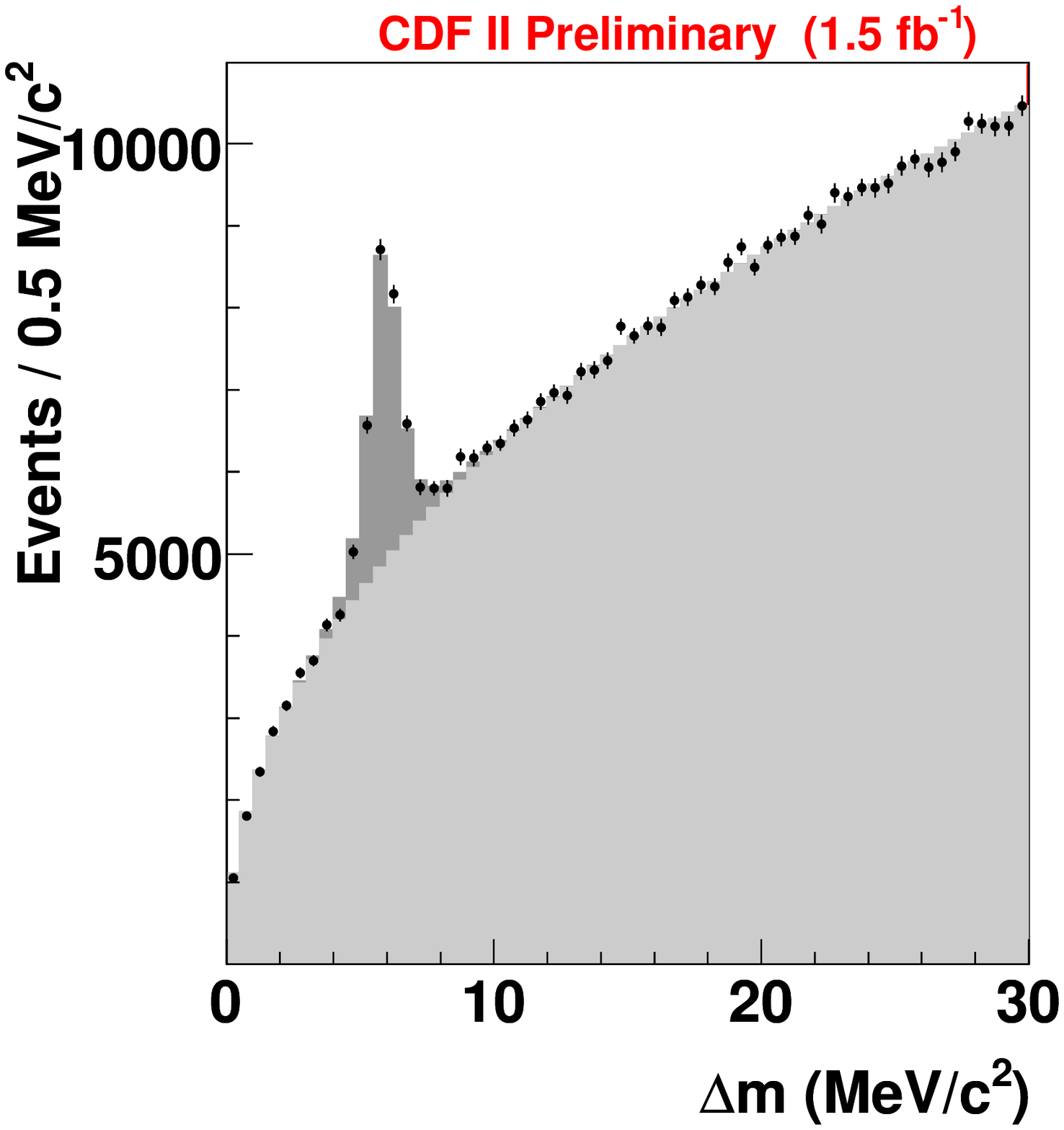}
}
\caption{$\Delta m = m(K\pi\pi) - m(K\pi) - m(\pi)$ distributions
for Cabibbo-favored (left) and doubly Cabibbo-suppressed candidates (right).  
The combinatorial background and signal are shown in light and dark 
grey respectively.} \label{fig:dmixdm}
\end{figure*}

The analysis~\cite{:2007uc} looks for a mixing-induced time dependence of the ratio of 
$(D^0 \to K^+ \pi^-)/(D^0 \to K^- \pi^+)$ decays in a sample of events collected
using CDF's impact parameter trigger.  In this case, the flavor of the $D^0$
meson at production is tagged by selecting the decay 
$D^{*+} \to \pi^+ D^0, D^0 \to K\pi$ chain.  To reduce backgrounds, particle 
identification (PID) is applied as in the $B_s$ mixing analysis, and $K\pi$ 
candidates that fall into the $D^0$ mass window with both mass assignments
to the reconstructed tracks are rejected (PID separation has limited 
discriminating power).  Events are then selected using the 
$\Delta m = m(K\pi\pi) - m(K\pi) - m(\pi)$ distributions shown in 
Fig.~\ref{fig:dmixdm}.

\begin{figure*}[t]
\centering
\subfigure[]{
\includegraphics[width=0.35\textwidth]{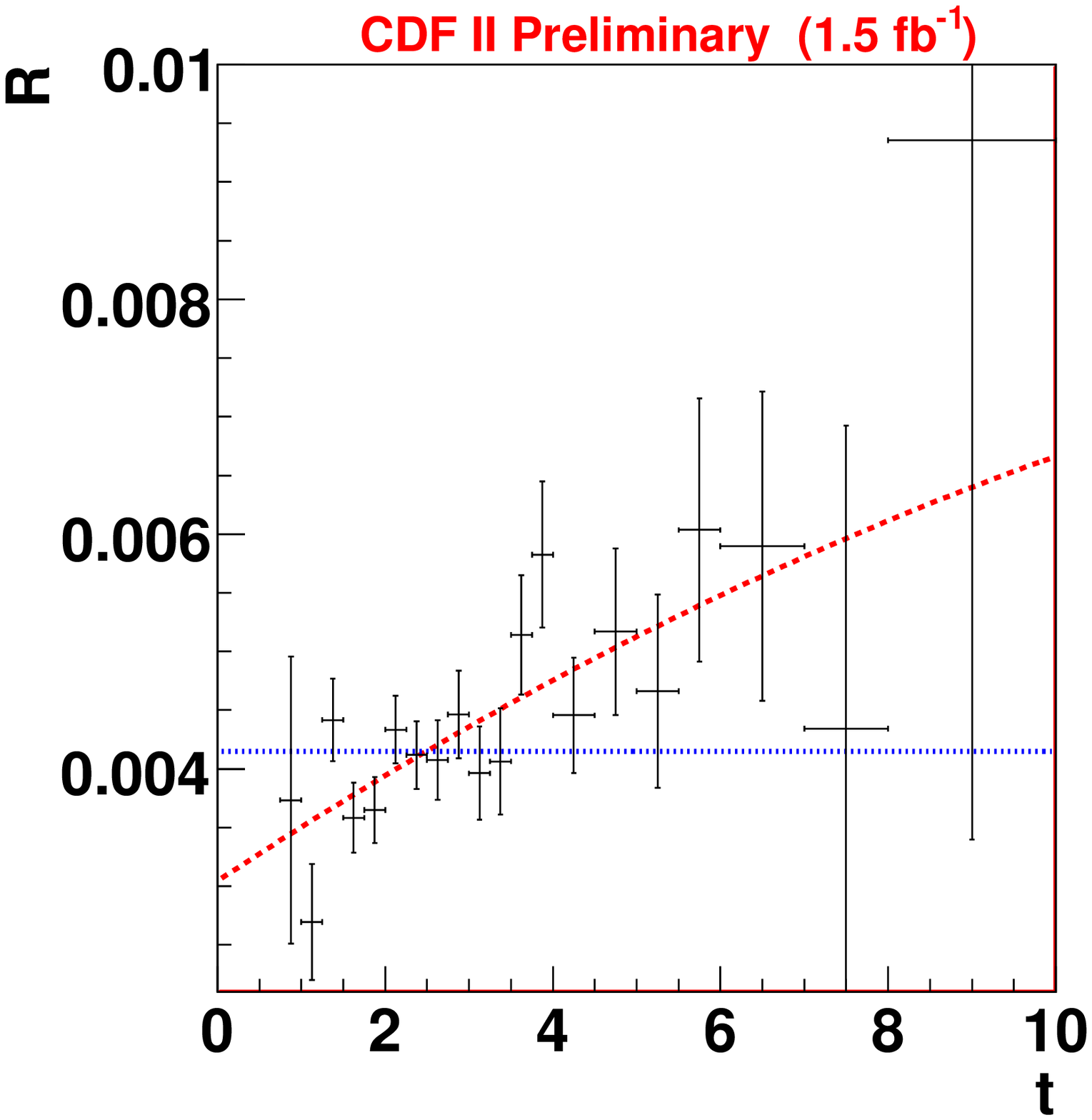}
}
\subfigure[]{
\includegraphics[width=0.35\textwidth]{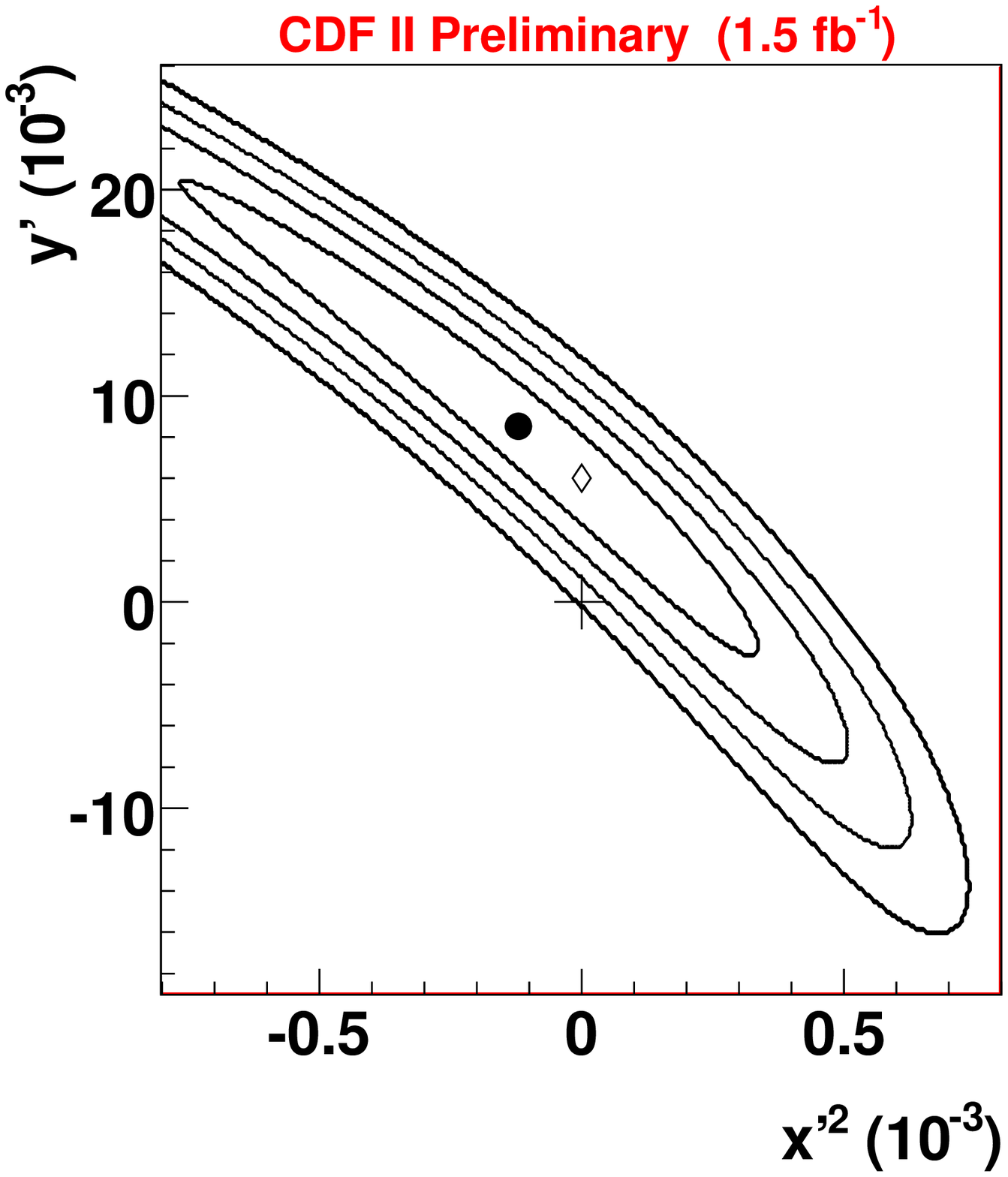}
}
\caption{Ratio of doubly Cabibbo-suppressed over Cabibbo-favored decays as a function of
proper decay time (left), and (bayesian) probability intervals for the mixing parameters 
($x'^2,y'$) (right).} \label{fig:dmixr}
\end{figure*}

Figure~\ref{fig:dmixr} shows the results of the analysis.  On the left, the ratio of 
doubly Cabibbo-suppressed over Cabibbo-favored decays as a function of
proper decay time exhibits the parabolic dependence
\begin{equation} 
R(t) = R_D + \sqrt{R_D}y't + \frac{x'^2 + y'^2}{4}t^2
\end{equation}
expected in the presence of mixing, where $x',y'$ are the mixing 
parameters defined in section~\ref{sec:mixing}.
The $\chi^2/dof$ for the parabolic (flat) fit is 19.2/17 (36.8/19)
The right plot shows the probability intervals in the ($x'^2,y'$) plane.  
The best fit is 3.8$\sigma$ away
from the null hypothesis.

\section{CP VIOLATION \label{sec:cpv}}
In this paper, CP violation measurements are split in three categories:
\begin{itemize}
\item CP violation in the decay amplitudes, where 
$\Gamma(B \to f) \neq \Gamma(CP(B) \to CP(f))$.  In charged meson decays, this is the 
only possible manifestation of CP violation, and a measurement in the $B^+ \to J/\psi K^+$
channel is described.
\item CP violation in mixing, which can be detected through an asymmetry in charged-current
neutral meson decays.  Recent results from inclusive same-sign dimuon events and exclusive
$B_s \to \mu^+ \nu D_s^-X$ decays are given.
\item CP violation in the interference between mixing and decay can be tested in decays
with and without mixing to a single final state.  Here, studies in the $B_s \to J/\psi \phi$
channel are presented.
\end{itemize}

\subsection{CP Violation in Decays \label{subsec:bplus}}
\begin{figure*}[t]
\centering
\subfigure[]{
\includegraphics[width=0.35\textwidth]{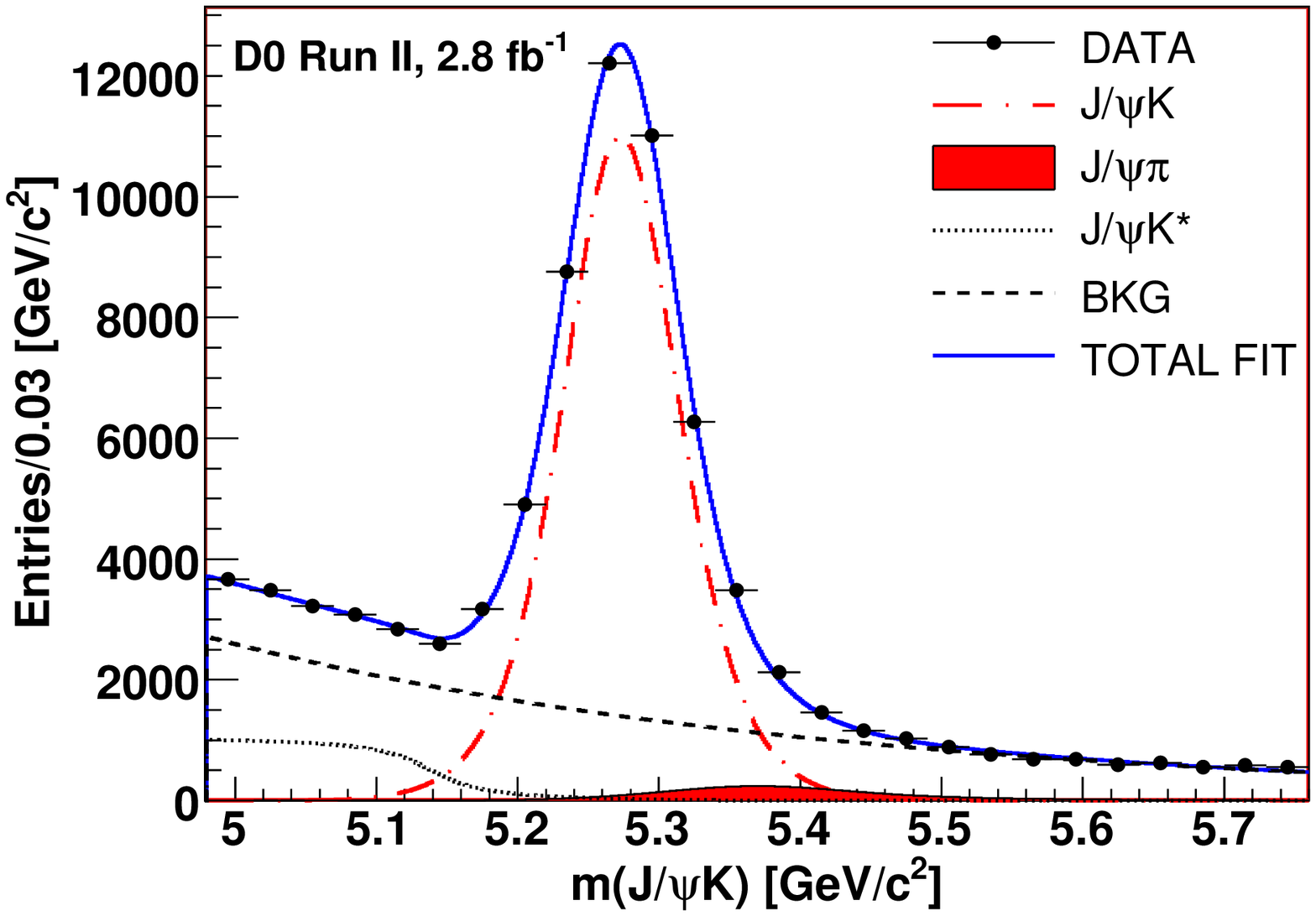}
}
\subfigure[]{
\includegraphics[width=0.35\textwidth]{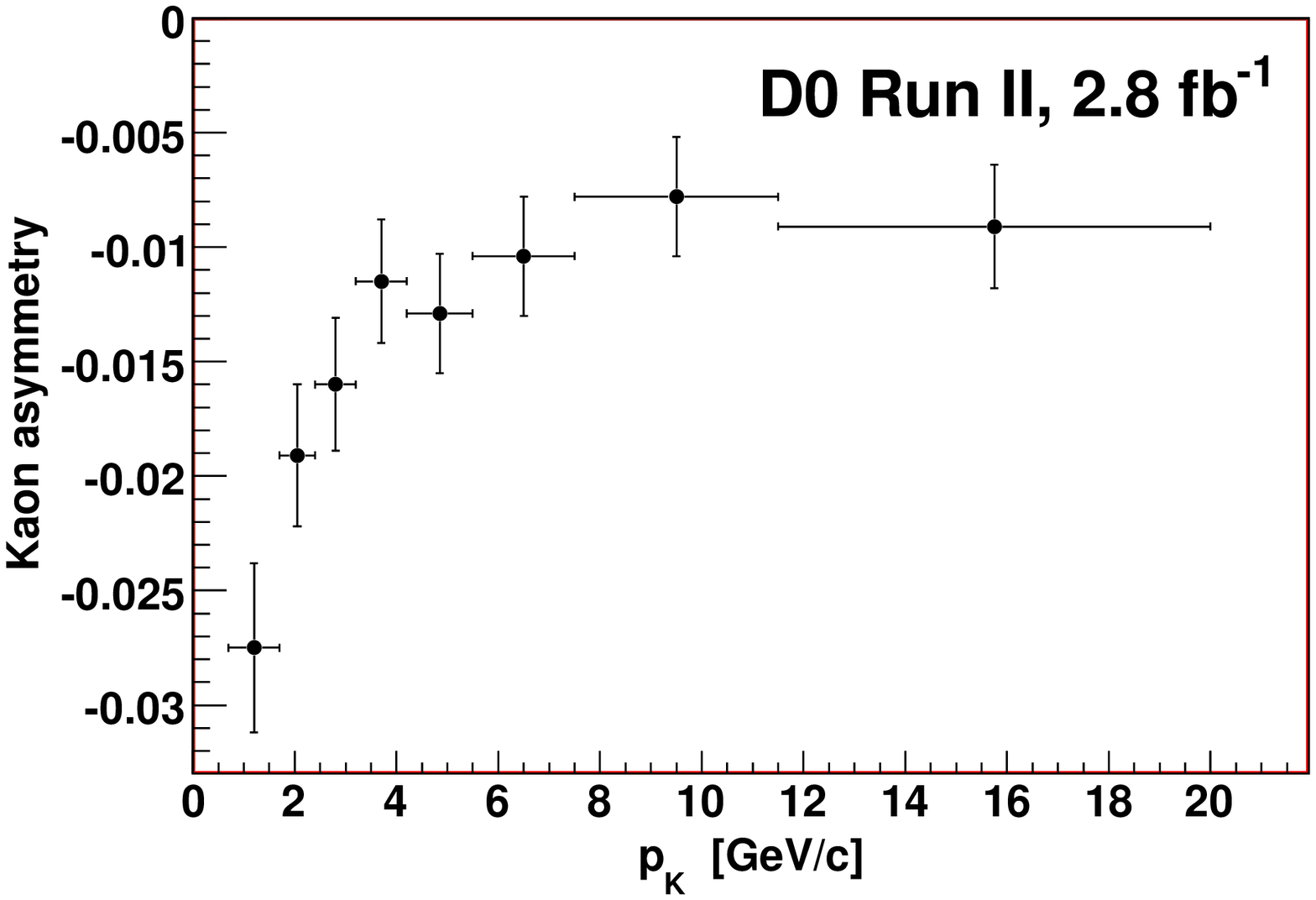}
} 
\caption{Left: $B^+$ candidate mass distribution showing the data and contributions extracted
from a fit to the data.  The shapes of the various contributions are extracted from the simulation.
Right: kaon reconstruction asymmetry as a function of kaon momentum.} \label{fig:bplus}
\end{figure*}

D0 selects a $B^+ \to J/\psi K^+ (\pi^+)$ sample~\cite{Abazov:2008gs} 
where the $J/\psi$ decays to a pair of 
muons.  This yields a very clean sample with approximately 
40,000 candidates as shown in the 
left panel of Fig.~\ref{fig:bplus}.  To determine the 
asymmetry, the sample is divided in 8 subsamples based on solenoid polarity $\beta$,
sign $\gamma$ of the $J/\psi K^+$ system pseudo-rapidity $\eta$, and kaon charge $q$:
\begin{equation}
n_q^{\beta\gamma} = \frac{1}{4}N\epsilon^\beta(1+qA)(1+q\gamma A_{fb})
(1+\gamma A_{det})(1+q\beta\gamma A_{ro})(1+q\beta A_{q\beta})
(1+\beta\gamma A_{\beta\gamma}),
\end{equation}
where N is the total number of events in the sample, $\epsilon^\beta$ is the fraction of 
integrated luminosity taken with solenoid
polarity $\beta$, $A$ is the CP-violating asymmetry, $A_{fb}$ is the forward-backward asymmetry
and all other asymmetries are potentially generated by the detector.  
The total number of events in the $J/\psi K^+$ and $J/\psi \pi^+$ channels, the fraction 
of integrated luminosity for each solenoid polarity and the asymmetries are determined
from a fit using the number of events in each subsample for inputs.  The forward-backward
symmetry, as well as all detector asymmetries are found to be compatible with zero.  At 
this stage, the observed CP asymmetry needs to be corrected for the asymmetry in kaon 
reconstruction efficiency induced by the different interactions of $K^+$ and $K^-$ with
the detector, which is made of matter only.  This reconstruction efficiency asymmetry is 
measured from a sample of $D^{*+} \to D^0\pi^+, D^0 \to \mu \nu K^-$ events as a function
of kaon momentum (right panel of Fig.~\ref{fig:bplus}), and applied to the kaon spectrum in
the $B^+$ sample.  After this correction, the CP asymmetries are measured to be
$A_{CP}(B^+ \to J/\psi K^+) = 0.0075 \pm 0.0061 (stat) \pm 0.0027 (syst)$ and
$A_{CP}(B^+ \to J/\psi \pi^+) = -0.09 \pm 0.08 (stat) \pm 0.03 (syst)$.  The dominant
systematic uncertainty comes from the mass distribution model.

\subsection{CP Violation in Mixing \label{subsec:semil}}
CP violation due to mixing between neutral $B$ mesons can be searched for by measuring
the asymmetry in same-sign dimuon events:
\begin{equation}
A_{SL}^{\mu\mu} = \frac{N(b\bar{b} \to \mu^+ \mu^+ X) - N(b\bar{b} \to \mu^- \mu^- X)}
                       {N(b\bar{b} \to \mu^+ \mu^+ X) + N(b\bar{b} \to \mu^- \mu^- X)}.
\end{equation}
Two approaches are used:
\begin{itemize}
\item CDF~\cite{cdfdimu} uses all dimuon events but exploits the two-dimensional 
($\mu-\mu$) impact parameter significance
distributions to unfold contributions from different sources, including beauty and
charm hadrons.  CDF finds 
$A_{SL}^{\mu\mu} = -0.0080 \pm 0.0090 (stat) \pm 0.0068 (syst)$.
\item D0~\cite{Abazov:2006qw} uses all dimuon events, 
and estimates the contributions from all possible 
processes to the sample, including sequential decays, Drell-Yan, instrumentals, etc.
The ``8 subsamples'' technique described in section~\ref{subsec:bplus} is used to 
extract the asymmetry.  D0 measures 
$A_{SL}^{\mu\mu} = -0.0053 \pm 0.0025 (stat) \pm 0.0018 (syst)$.
\end{itemize}
In both cases, the asymmetry from fake muons due to kaons and pions which have 
asymmetric reconstruction efficiencies (see section~\ref{subsec:bplus}) is 
determined from data and corrected for.

The flavor-specific asymmetry $A_{SL}^s$ then needs to be derived using the 
sample composition and the known oscillation probabilities, since $A_{SL}^{\mu\mu}$
has contributions from both $B_d$ and $B_s$ mixing.  This is done by taking $A^d_{SL}$
from the $B$-factories, and the known $B_d$ and $B_s$ production ratios and mixing 
parameters.  The results are:
\begin{itemize}
\item CDF: $A_{SL}^s = 0.020 \pm 0.021 (stat) \pm 0.016 (syst) \pm 0.009 (inputs)$;
\item D0: $A_{SL}^s = -0.0064 \pm 0.0101 (all\ uncertainties\ combined)$.
\end{itemize}
An alternative
approach is to use flavor-specific decays, as is done in the D0 
analysis~\cite{Abazov:2007nw} of $B_s \to \mu^+ \nu D_s^- X$, where again
the asymmetry is determined using the 8 subsamples technique.  This yields
$A_{SL}^s = 0.0245 \pm 0.0193 (stat) \pm 0.0035 (syst)$.

\subsection{CP Violation in the Interference Between Mixing and Decay \label{subsec:phi}}
The search for CP violation in the interference of mixing and decay can be searched for
in final states that are common to both $B_s$ and $\bar{B}_s$ decays.  Both 
CDF~\cite{Aaltonen:2007he} and D0~\cite{:2008fj} make flavor-tagged
measurements of the CP properties of the $B_s (\bar{B}_s) \to J/\psi \phi$ process.  This is
similar to the measurement of $\sin (2\beta)$ in $B_d (\bar{B}_d) \to J/\psi K_s$
decays at the $B$-factories.  The angle $\phi_s (= -2\beta_s)$ is however much smaller.
An easy way to see this is that the opposing side in the corresponding unitarity triangle
is proportional to $\left|\frac{V_{us}V_{ub}^*}{V_{cs}V_{cb}^*}\right|$, whereas for 
$\beta$ it is proportional to $\left|\frac{V_{ud}V_{ub}^*}{V_{cd}V_{cb}^*}\right|$.
This measurement requires the determination of the CP eigenvalue
of the vector-vector $J/\psi \phi$ final state, 
and benefits from the identification of the flavor of the 
decaying $B_s$ meson.

\begin{figure*}[t]
\centering
\subfigure[]{
\includegraphics[width=0.40\textwidth]{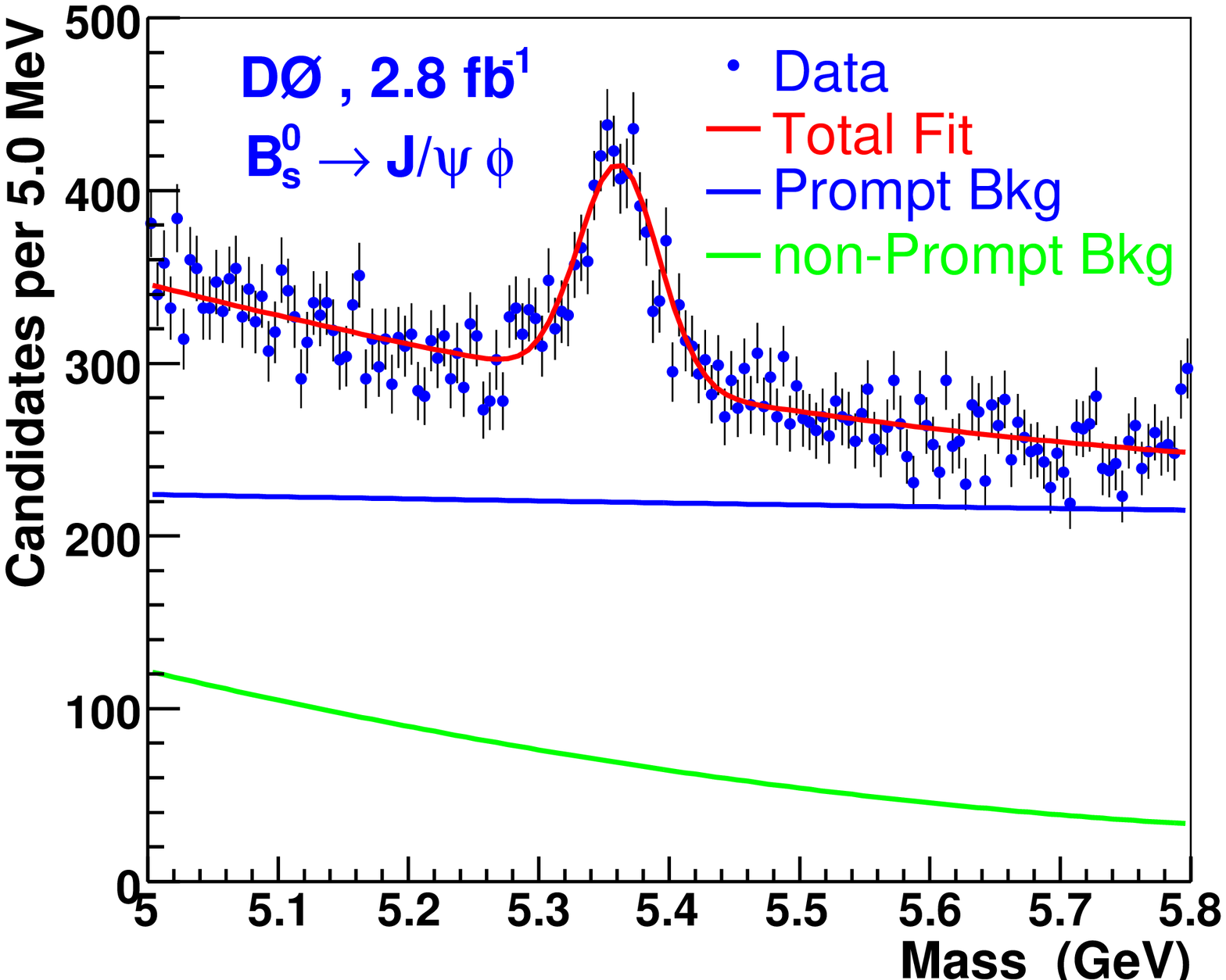}
} 
\subfigure[]{
\includegraphics[width=0.35\textwidth]{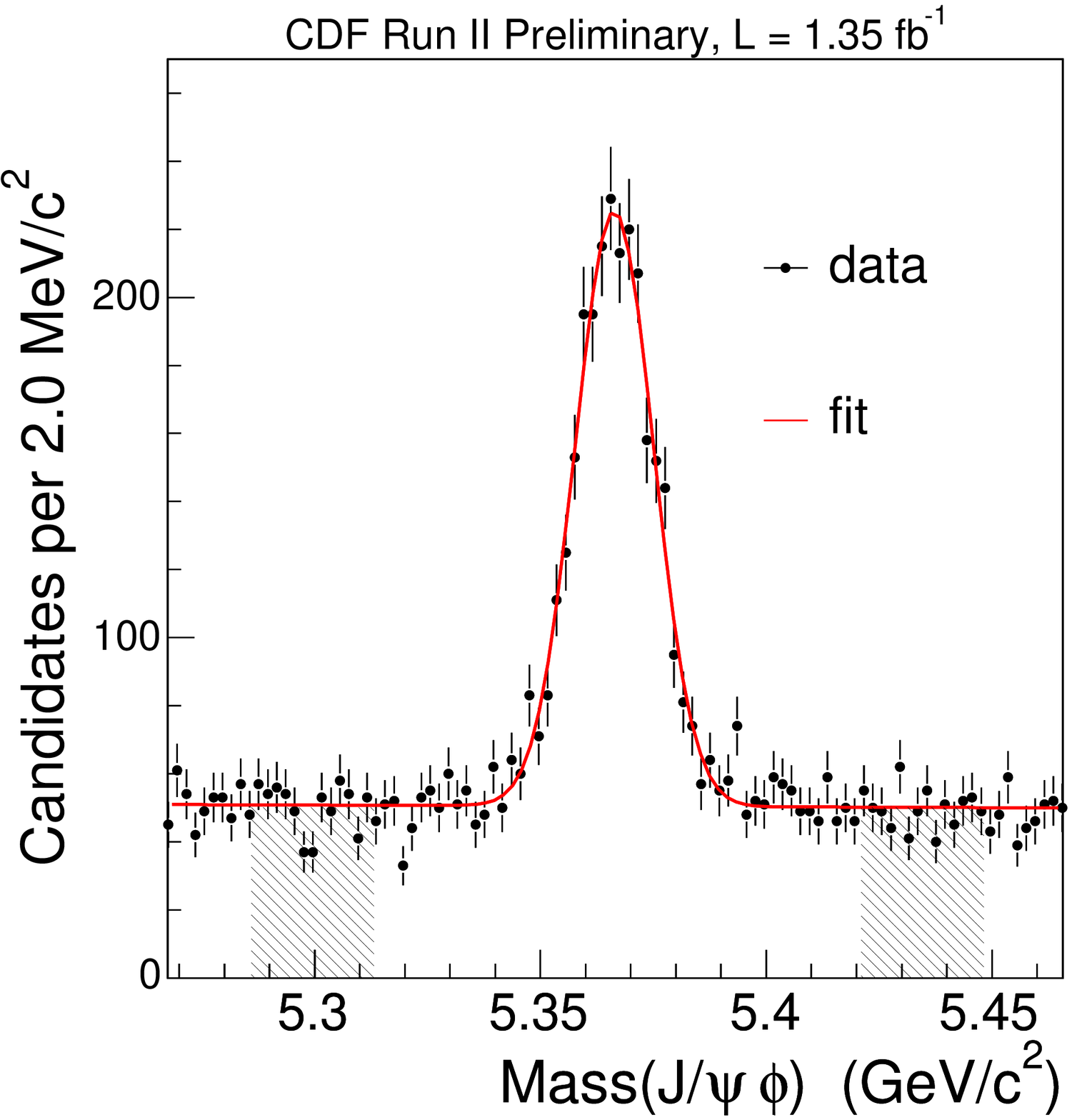}
}
\caption{$B_s \to J/\psi \phi$ candidate samples after event selection in 
D0 (left) and CDF (right).} \label{fig:phisample}
\end{figure*}

Both experiments use events where the $J/\psi$ meson decays to a pair of muons, leading
to a high trigger and reconstruction efficiency.  At the event selection level, CDF
uses a neural network with variables that include PID from $dE/dx$ in the drift chamber
and information from the time-of-flight detector.  D0 uses a ``square cuts'' event selection 
without PID.  The resulting $B_s$ meson candidate samples are shown in 
Fig.~\ref{fig:phisample}.

Since the $J/\psi \phi$ final state consists of two vector mesons, both CP eigenvalues 
are possible.  The value for a given $B_s$ candidate is measured by determining
the relative polarization of the $J/\psi$ and $\phi$ mesons: the angular dependence
of the relative directions of the decay products $\mu^+ \mu^-$ and $K^+ K^-$ is 
expressed in the $J/\psi$ restframe (``transversity basis'') in terms of this 
polarization.  The corresponding probability density functions are determined from 
simulation, and used to determine each candidate's probability to be in a given 
eigenstate.  Note that the detector efficiencies are not flat as a function of 
transversity angles, and this is properly accounted for.

Constraining the flavor of the decaying $B_s$ meson removes a twofold ambiguity 
in the result.  This is achieved by determining the $B_s$ meson flavor at production
and using the mixing frequency measured by CDF (see section~\ref{subsec:bsmix}).
Both experiments use both opposite-side and same-side tags as described in 
section~\ref{subsec:bsmix}.  The tagging power
obtained in CDF is $\epsilon {\cal D}^2$ = 1.28\% and 3.65\% for opposite-
and same-side tags respectively, and 4.68\% for the combined tag in D0.

To maximize sensitivity, an unbinned likelihood fit is used which assigns event-by-event
signal and background probabilities based on reconstruction uncertainties, dilution,
accuracy of CP eigenstate determination etc.  D0 uses constraints on strong phases
from world-average values~\cite{Barberio:2007cr} of measurements of
$B_d \to J/\psi K^*$ decays, whereas CDF lets these phases float.  
Constraining the phases essentially eliminates the remaining ambiguity in 
the extraction of the result, but there may be differences between the phases in 
$B_d \to J/\psi K^*$ and $B_s \to J/\psi \phi$ decays.  From the likelihood fit,
the average lifetime, $\Delta \Gamma_s$, $\phi_s (= -2\beta_s)$, the magnitudes
of the polarization amplitudes and the strong phases are extracted.

\begin{figure*}[t]
\centering
\subfigure[]{
\includegraphics[width=0.42\textwidth]{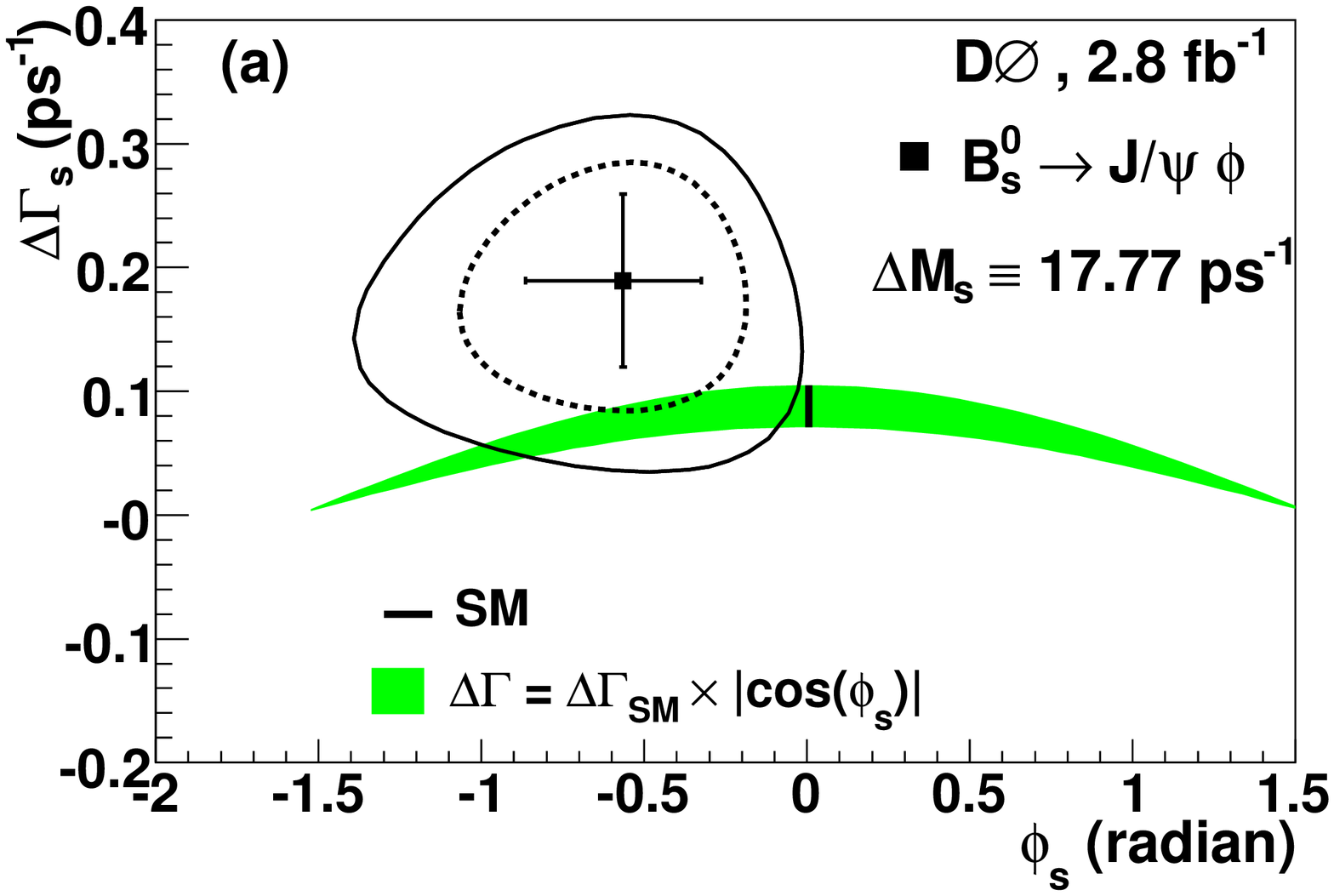}
} 
\subfigure[]{
\includegraphics[width=0.35\textwidth]{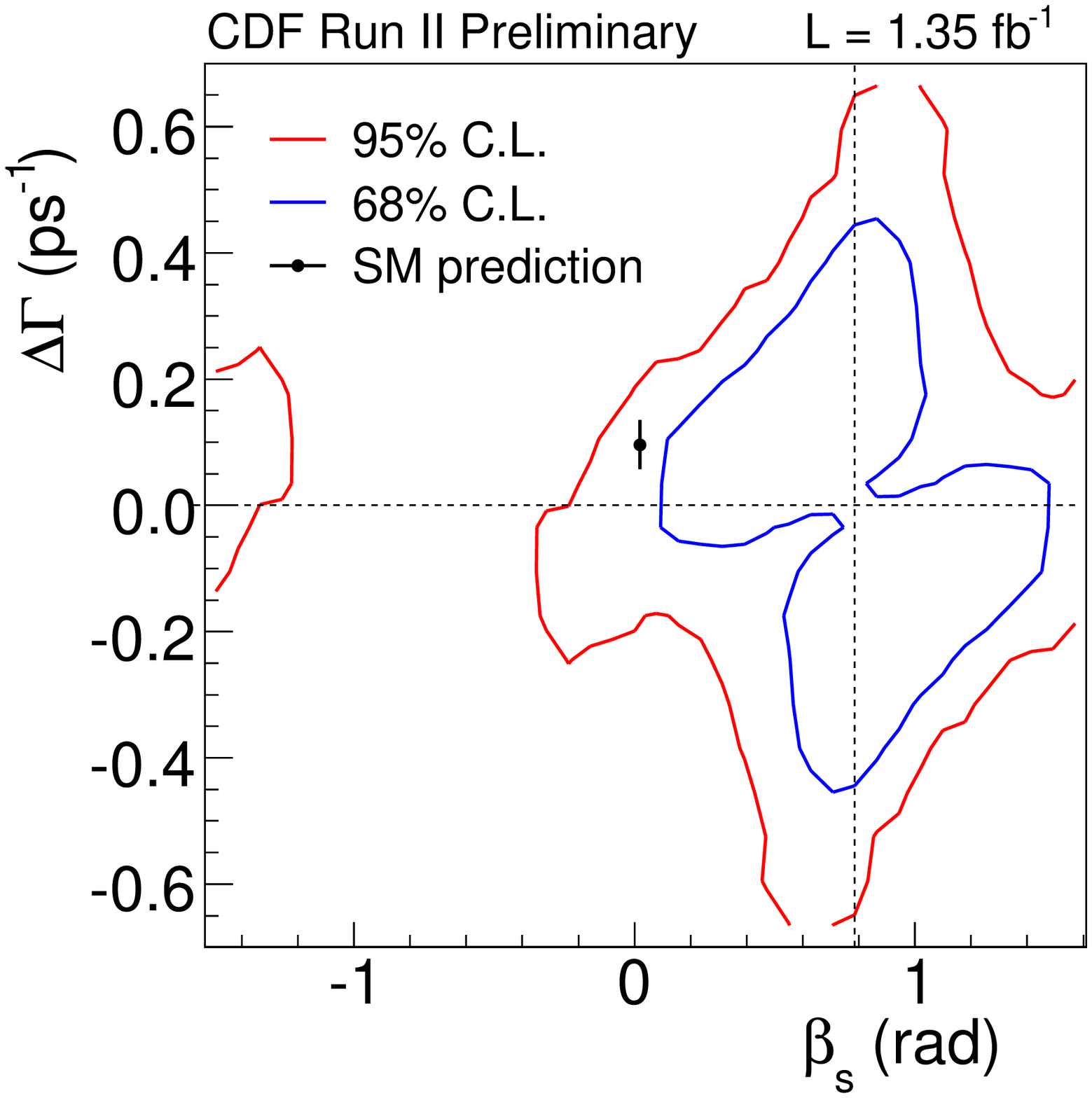}
}
\caption{Two-dimensional confidence intervals in the ($\beta_s$,$\Delta \Gamma_s$) plane   
from the flavor-tagged analysis of $B_s \to J/\psi \phi$ decays from
D0 (left) and CDF (right).} \label{fig:phires}
\end{figure*}

A small added complication to the extraction of the final CDF result is due to the
fact that in that analysis the likelihood profiles are not parabolic close to 
the minima.  Since standard frequentist techniques can't be applied, CDF uses a 
Feldman-Cousins-like likelihood ratio ordering to build the two-dimensional 
confidence region in the ($\beta_s$,$\Delta \Gamma_s$) plane.  The results 
obtained by CDF and D0 are shown in Fig.~\ref{fig:phires}.

Both experiments measure values that are in reasonable agreement with the standard model:
the ``standard model probability'' is 15\% and 7\% for the CDF and D0 results respectively.
However, it is clear from Fig.~\ref{fig:phires} that both results pull in the same direction
($\phi_s (= -2\beta_s)$) and with similar magnitudes.  While an official combination of
these results by the experiments is not ready yet, the UT{\em fit} collaboration has 
performed a combined fit of these results~\cite{Bona:2008jn} with the
CP asymmetries described in section~\ref{subsec:semil}.  They try multiple methods to 
unfold the strong phases constraint from the D0 result, and in each of the methods they
find a value of $\phi_s$ that is 3$\sigma$ or more away from the standard model prediction.
It should be noted that for both CDF and D0, the measurement uncertainties are completely
dominated by statistics, so that the results will still improve by quite a bit, hopefully
leading to a conclusive statement.  If the UT{\em fit} conclusion is confirmed (and 
strengthened), this would be the \underline{first evidence for CP violation outside the 
CKM mechanism}.

\section{CONCLUSIONS \label{sec:concl}}

The beauty and charm physics programs at CDF and D0 continue to produce a large 
number of excellent results.  A number of recent highlights were presented here.
CDF has now seen evidence for $D$ meson mixing, and both CDF and D0 are making 
precise measurements of CP violation asymmetries in $B$ meson mixing and decay.
The $B_s$ meson mixing parameters are being pinned down: CDF measures $\Delta m_s$
with high precision, both experiments find a value of $\Delta \Gamma_s$ in good
agreement with the standard model, and both experiments see similar deviations
of $\phi_s$ from the standard model expectation.  The latter measurement will
become significantly more precise with increasing statistics.
If the result is confirmed, this will be
the first sign of CP violation in the quark sector outside the CKM mechanism.

\begin{acknowledgments}
We thank the staffs at Fermilab and collaborating institutions, 
and acknowledge support from the 
DOE and NSF (USA);
the A.P. Sloan Foundation (USA); 
the INFN (Italy); 
the Ministry of Education, Culture, Sports, Science and Technology (Japan); 
the Swiss National Science Foundation (Switzerland); 
the CICT (Spain); 
the European Community's Human Potential Programme under contract HPRN-CT-2002-00292; 
the Academy of Finland (Finland);
CEA and CNRS/IN2P3 (France);
FASI, Rosatom and RFBR (Russia);
CNPq, FAPERJ, FAPESP and FUNDUNESP (Brazil);
DAE and DST (India);
Colciencias (Colombia);
CONACyT (Mexico);
KRF and KOSEF (Korea);
CONICET and UBACyT (Argentina);
FOM (The Netherlands);
STFC (United Kingdom);
MSMT and GACR (Czech Republic);
CRC Program, CFI, NSERC and WestGrid Project (Canada);
BMBF and DFG (Germany);
SFI (Ireland);
The Swedish Research Council (Sweden);
CAS and CNSF (China);
and the
Alexander von Humboldt Foundation (Germany).
\end{acknowledgments}

\bibliographystyle{utphys}
\bibliography{procs}




\end{document}